\documentclass[
tightenlines,
nofootinbib,
twocolumn,
 amsmath,amssymb,
 aps,
]{revtex4-1}

\usepackage{color,graphicx}
\usepackage{tcolorbox}
\usepackage{dcolumn}
\usepackage{bm}
\usepackage{slashed}
\usepackage{cancel}
\usepackage{mathtools}
\usepackage{soul}
\begin{document}

\title{Is the Momentum Sum Rule Valid for Nuclear Structure Functions ?}

\author{Stanley J. Brodsky}
\affiliation{SLAC National Accelerator Laboratory, Stanford University}

\author{Ivan Schmidt}
\affiliation{Departamento de F\'\i sica y Centro Cient\'\i fico
Tecnol\'ogico de Valpara\'\i so-CCTVal \\ Universidad T\'ecnica
Federico Santa Mar\'\i a, Casilla 110-V, Valpara\'\i so, Chile}

\author{Simonetta Liuti}%
\altaffiliation[Also at ]{Laboratori Nazionali di Frascati, INFN, Frascati, Italy}
\email{sl4y@virginia.edu}
\affiliation{%
 Department of Physics, University of Virginia, Charlottesville, VA 22904, USA. 
}%

\date{\today}

\allowdisplaybreaks

\begin{abstract}
We address the validity of the momentum sum rule for deep inelastic nuclear structure functions. 
\end{abstract}

\maketitle

\begin{figure}
\begin{center}
\includegraphics[scale=0.4]{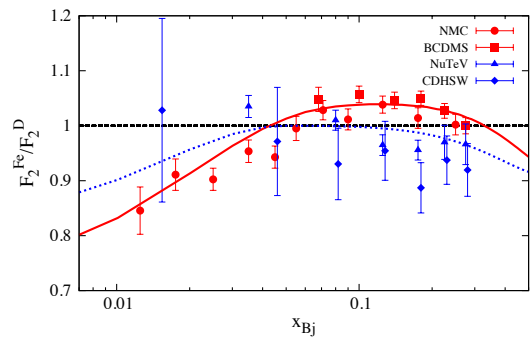}
\vspace{-5cm}
\caption{Comparison of the ratio of iron to deuteron nuclear structure functions measured in deep inelastic neutrino-nucleus scattering (NuTeV \cite{Tzanov:2005kr}, CDHSW \cite{Berge:1989hr}), and muon-nucleus scattering (BCDMS \cite{Benvenuti:1987az} and NMC \cite{Amaudruz:1995tq,Arneodo:1996rv}).
All data are displayed in the online Durham HepData Project
Database \cite{HepData}. 
Anti-shadowing is absent in the neutrino charged current data.}
\label{Comparison}
\end{center}
\end{figure}

One of the most surprising results of the NuTeV measurement~\cite{Tzanov:2005kr} of nuclear structure functions in deep inelastic charged-current reactions $\nu A \to \mu X$ is the absence of anti-shadowing in the domain $0.1 < x_{Bj}< 0.2$. 
As shown by Refs.~\cite{Schienbein:2007fs,Kovarik:2011dr,Kalantarians:2017mkj}, the parton distribution of nuclei measured in deep inelastic neutrino reactions  shows no enhancement above additivity in this domain, and is thus distinctly different from the corresponding Parton Distribution Function (PDF) measured in  charged lepton deep inelastic scattering (Figure \ref{Comparison}). 
This striking difference between neutrino vs.  deep inelastic scattering  measurements is in direct conflict with the conventional expectation that the quark and gluon distributions of the nucleus are universal properties of the nuclear eigenstate and are thus process independent.  Moreover, the NuTeV measurement contradicts the expectation that anti-shadowing,  $\frac{\sigma_A}{ A \sigma_N} > 1$, will compensate  for the shadowing  of the nuclear PDF,
$\frac{\sigma_A}{A \sigma_N} < 1 $, which is observed in the domain $0.1 < x_{Bj}< 0.2$  in order to restore the momentum sum rule~\cite{Nikolaev:1975vy,Frankfurt:1990xz}.

%

%
%

The NuTeV results thus call into question theoretical expectations concerning the fundamental nature of leading-twist deep inelastic scattering reactions on nuclei.

The shadowing vs. anti-shadowing of nuclear cross sections can be understood as a Glauber  phenomenon involving the constructive vs. destructive interference of two-step 
and one-step amplitudes illustrated in Figure \ref{OneandTwoStep} \cite{Stodolsky:1994ka}.
The key points underlying this mechanism are:

\begin{figure}
\begin{center}
\includegraphics[scale=0.22]{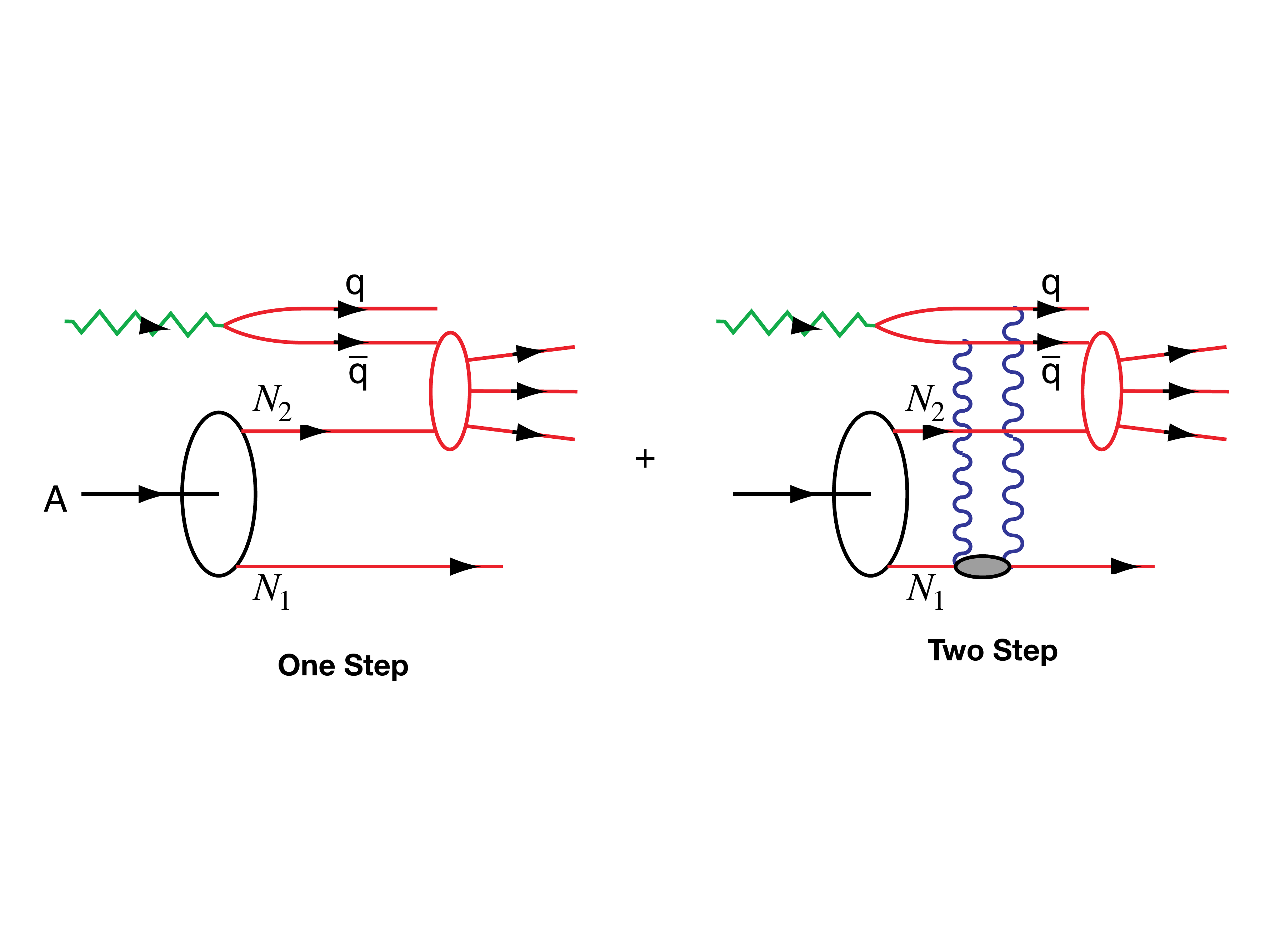}
\vspace{-2cm}
\caption{Sum of interfering one-step and two-step amplitudes in DIS on a nucleus 
$A$:  $\gamma^* A \to X. $ 
The initial scattering in the two step amplitude on the front-face nucleon $N_1$ is diffractive DIS:
$\gamma^* N_1 \to [q \bar q] N_1^\prime$ which leaves $N_1$ intact. The propagating vector $[q \bar q]$ system then interacts inelastically on $N_2$:  
               $ [q \bar q]+ N_2 \to X$. 
               The two step amplitude interferes with the one-step amplitude $ \gamma^*+ N_2 \to X$ on $N_2$.  The interior nucleon $N_2$ sees two fluxes, the virtual photon $\gamma^*$ and the secondary beam $[q\bar q ]$ generated by DDIS on $N_1$.   In effect, nucleon $N_1 $ ``shadows" $N_2$.}
\label{OneandTwoStep}
\end{center}
\end{figure}

\begin{figure}
\begin{center}
\includegraphics[scale=0.25]{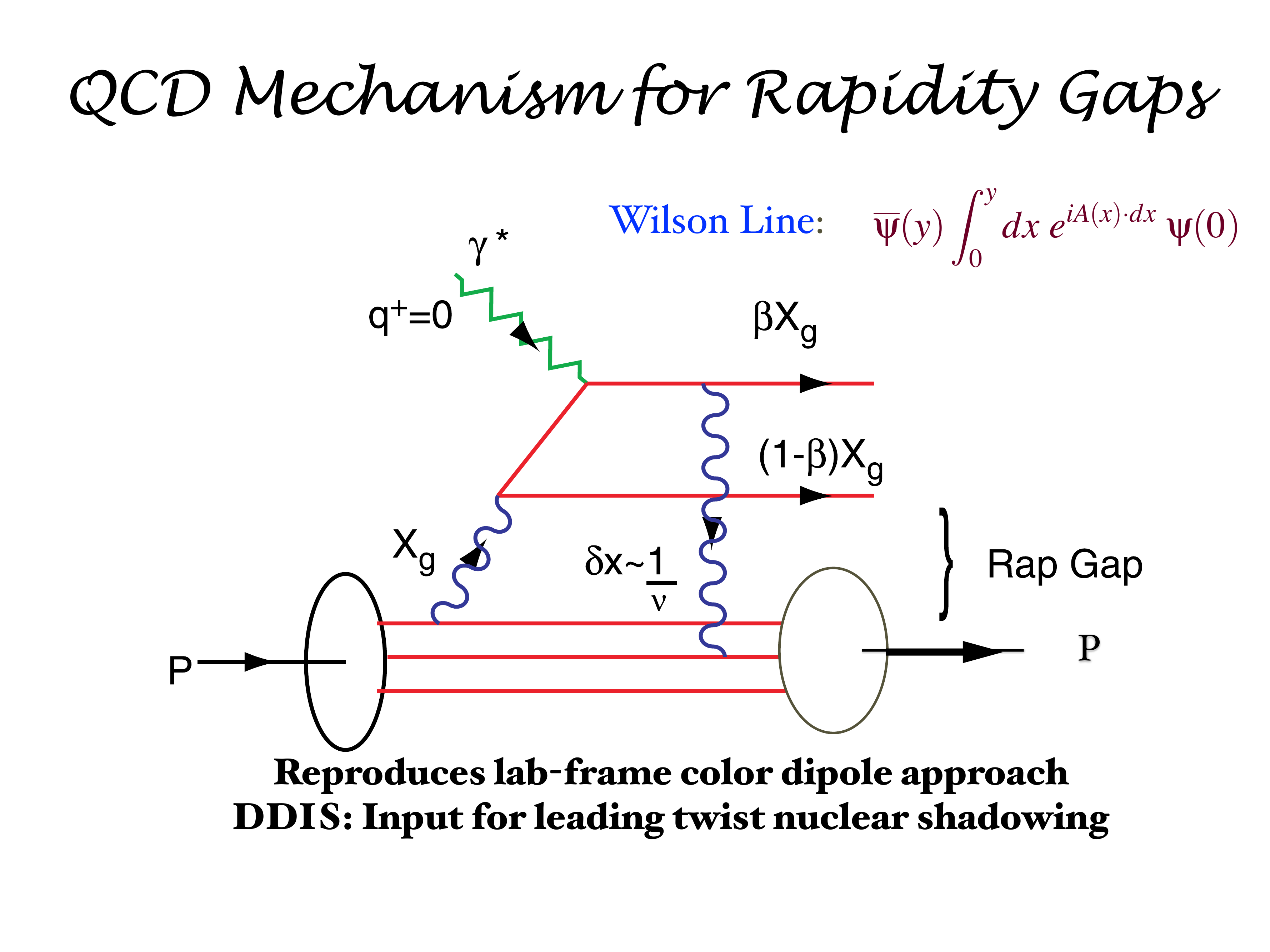}
\caption{QCD mechanism for leading-twist diffractive DIS $\gamma^* p \to p' X$. The second gluon exchanged between the $[q\bar q]$ and the spectator quark of the proton occurs after the quark has been struck by the lepton as in a Wilson loop. The two gluons in the t-channel correspond to color-singlet Pomeron exchange.  Since the intermediate state can propagate on-shell it gives a Glauber cut and phase $i$. A similar final state interaction also leads to the Sivers effect, the $i\vec S_p \cdot \vec q \times \vec p_q$ pseudo-time-odd correlation.}
\label{DDIS}
\end{center}
\end{figure}
%
\begin{figure}
\begin{center}
\includegraphics[scale=0.2]{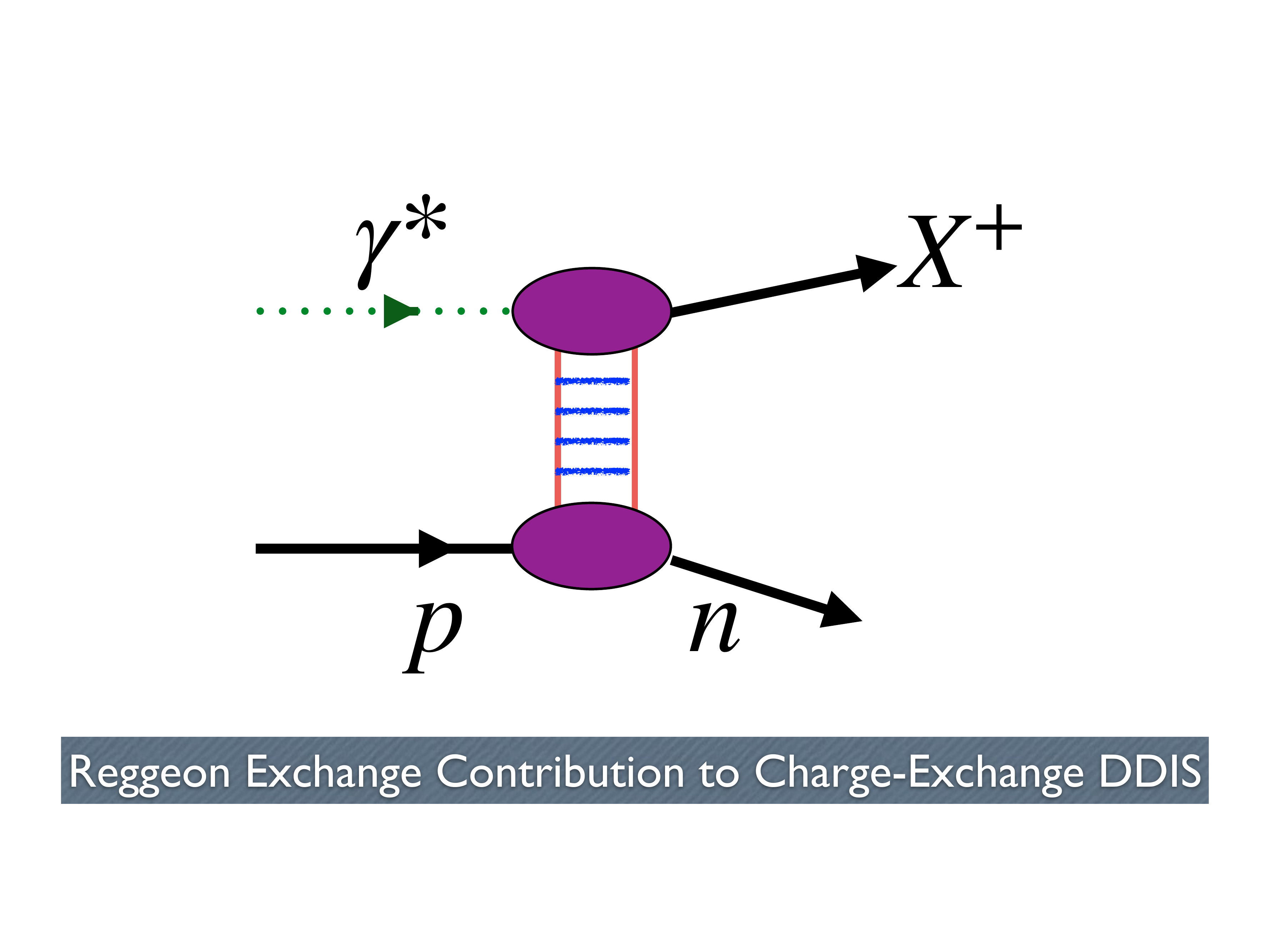}
\caption{QCD mechanism for charge-exchange leading-twist diffractive DIS $\gamma^* p \to n X^+$. }
\label{CEDDIS}
\end{center}
\end{figure}

\begin{itemize}
\item The first step of the two-step amplitude involves leading-twist  Diffractive Deep Inelastic Scattering  (DDIS) on a front-face nucleon $N_1$ which leaves the nucleon intact.  In fact, DDIS in $\gamma^* N \to N X$ reactions has been observed to satisfy Bjorken scaling, and approximately 10\% of high energy DIS events are diffractive \cite{Derrick:1993xh,Ahmed:1994nw}.  

\item The second step of the two-step amplitude involves the inelastic scattering of the state $X$ on a second nucleon $N_2$. The interference of the two-step amplitude with the standard DIS event on nucleon $N_2$ can produce shadowing or anti-shadowing of the nuclear PDF depending on the phase of the DDIS amplitude.  

\item In the Regge theory of strong interactions diffraction occurs  through the exchange of either a Pomeron or a Reggeon trajectory.
In Quantum Chromodynamics (QCD), the Pomeron and the Reggeon correspond to two gluon and to quark-antiquark color-singlet exchanges, respectively.
The diffractive process is leading twist and it, therefore, displays Bjorken scaling.  
The leading-twist QCD mechanism that underlies DDIS is illustrated in Fig.\ref{DDIS}.

\item The occurrence of either shadowing or antishadowing is governed by the difference in the phase structure of Reggeon and Pomeron exchanges. The phase of the $ I = 0,1 $ Reggeon contributions to  DDIS  is  $\frac{1}{\sqrt 2}(-i + 1)$ with $\alpha_R = 1/2$. Its imaginary part 
is opposite to the positive
imaginary contribution of Pomeron exchange. When one multiplies by the phase $i$ from the propagating intermediate state, the relative phase of the two-step amplitude is thus destructive if DDIS is due to pomeron exchange (shadowing) or constructive  (anti-shadowing)  if the DDIS amplitude is due to Reggeon exchange.  The resulting effect  from the constructive interference appears in the 
$0.1 < x_{Bj} < 0.2$ domain of the nuclear PDF.
The exchange of the same Reggeon also leads to the Kuti-Weisskopf prediction: $F_{2}^p(x,Q^2) -  F_{2}^n(x,Q^2) \propto \sqrt{x}$ (this result is consistent with recent evaluations in Refs.\cite{Harland-Lang:2014zoa,Alekhin:2015cza}).

\item  Thus unlike shadowing, anti-shadowing from Reggeon exchange is flavor specific; {\it i.e.}, each quark and anti-quark will have distinctly different constructive interference patterns.  The flavor dependence of anti-shadowing explains why anti-shadowing is different for electron (neutral electromagnetic current) vs. neutrino (charged weak current) DIS reactions (Fig.\ref{Comparison}).

\item An important test of the explanation of anti-shadowing is to verify the existence of Bjorken-scaling  leading-twist charge exchange DDIS reaction  $\gamma^* p \to n X^+$ with a rapidity gap due to $I=1$ Reggeon exchange. Here $X^+$ is the sum of final states with charge $Q= 1$.
This process is shown in Figure \ref{CEDDIS}.
\end{itemize}

%
As a consequence of the Glauber processes with interfering amplitudes, the interior nucleons are shadowed at low $x_{Bj}$, while DIS at low $x_{Bj}$ occurs primarily on the front
nucleons.  This contradicts the OPE where the product
of currents acts uniformly on all quarks of the nucleus.
The interaction with a particular nucleon deep inside the nucleus depends on the survival of the projectile photon or its fluctuations reaching that nucleon.  Intuitively, one would expect that nucleon counting, which is a parton model way to understand the parton model sum rules, would fail. 
Technically, the derivation of sum rules which depends on both the operator product short distance and on the locality of  two currents 
in deeply virtual Compton scattering 
electron scattering at high $Q^2$, fails in a nucleus as we explain in detail in what follows. 

\begin{figure}
\begin{center}
\includegraphics[scale=0.35]{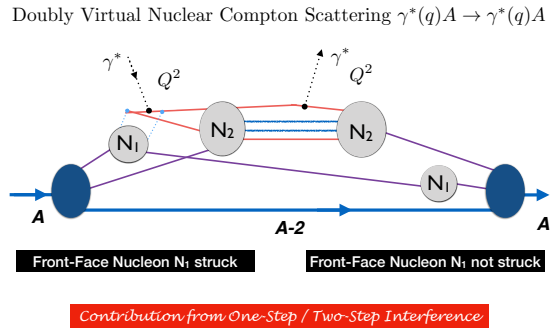}
\vspace{-2cm}
\caption{Contribution to doubly virtual Compton scattering on a nucleus $\gamma^* A \to \gamma^* A$ from the interference of the two-step and one-step amplitudes illustrated in Fig. \ref{OneandTwoStep}. The cut of the forward amplitude contributes to the inclusive DIS cross section $\gamma^* A \to X. $   This diagram cannot be reduced to a handbag amplitude where two currents interact on an interrupted quark propagator.}
\label{DVCS}
\end{center}
\end{figure}

The contribution to the forward virtual Compton scattering amplitude for a coherent \cite{Liuti:2005gi} scattering process on a nucleus,  $\gamma^*(Q^2) A \to \gamma^*(Q^2) A $, from the  interference between the two-step and one-step amplitudes
is shown in Fig. \ref{DVCS}.  Sum rules for deep inelastic scattering are analyzed using OPE for the forward virtual Compton scattering amplitude where
the moments of structure functions and other distributions can be evaluated as overlaps of the target hadron's light-front wavefunction
\cite{Brodsky:1980zm,Liuti:2013cna,Mondal:2015uha,Lorce:2011dv}.
The real phase of the resulting DIS amplitude and its OPE matrix elements reflects the real phase of the stable target hadron's wavefunction.

The usual ``handbag"  diagram where the two $J^\mu(x)$ and $J^\nu(0)$ currents acting on an uninterrupted quark  propagator  are replaced by a local  operator $ T^{\mu \nu}(0)$ as $Q^2 \to \infty$, is inapplicable in deeply virtual Compton scattering from a nucleus since the currents act on different nucleons. 

Unlike the handbag diagram, the phase of the deeply virtual amplitude arising from the Glauber interference amplitudes is always complex.  Thus the derivation of the momentum sum rule fails for the nuclear PDF: {\em  shadowing and anti-shadowing do not need to compensate each other to restore the momentum sum rule}.

Let us consider in more detail the interference of the single and the double step interactions shown in Fig.\ref{OneandTwoStep}.  Here $N_1$ is the front-face nucleon and $N_2$ an interior nucleon. In the one-step process only $N_2$ interacts via Pomeron exchange, while $N_1$ does not. In the two step, the scattering on $N_1$ is via Pomeron exchange, and both amplitudes have different phases, diminishing the $\bar{q}$ flux that reaches $N_2$. The interior nucleon is shadowed. The interior nucleon, $N_2$, thus sees two fluxes - the incident virtual photon $\gamma^*$ and the $V^0$ produced from DDIS on $N_1$. The relative phase of the one-step and two-step amplitudes is the critical factor of $i$ from the Glauber cut times the phase of Pomeron exchange in DDIS.   The destructive interference is why $N_2 $ does not see the full flux -- it is shadowed by $N_1$. 
Thus shadowing of the nuclear PDF is due to additional physical, causal events within the nucleus. 

Several additional points should be emphasized. First, the $q \bar{q}$ vector system $V^0$ propagates on-shell. 
This means that not all the propagators in the graph can be considered as being hard (of order $Q^2$): this invalidates the OPE, and, as a consequence, the momentum sum rule. 
Moreover, the finite path length due to the on-shell propagation of $V^0$  between $N_1$ and $ N_2$ contributes to the distance $({\Delta z})^2$ between the two virtual photons in the DVCS amplitude.  One no longer has $({\Delta z})^2 \approx {1/Q^2}$.  The distance between the currents cannot be less than the inter-nucleon distance, invalidating also the OPE and the parton momentum sum rule.
Finally, the interference diagram is real even if defined in the forward limit, since in the high $Q^2$ limit of the OPE only a local operator contributes, and the phases coming from Regge exchanges cancel each other. This certainly is another reason to invalidate the OPE for nuclei, since it would mean the absence of shadowing. 

Thus, the Glauber propagation of the vector system, $V^0$, produced by the  DDIS interaction on the nuclear front face and its subsequent  inelastic interaction with the nucleons in the nuclear interior $V^0 + N_b \to X$, occurs after the lepton interacts with the struck quark.  The corresponding amplitude for deeply virtual Compton scattering is not given by the handbag diagram alone since interactions between the two currents are essential.

Finally, we reiterate that because of the rescattering dynamics, the DDIS amplitude acquires a complex phase from Pomeron and Regge exchange;  thus final-state  rescattering corrections lead to  nontrivial ``dynamical" contributions to the measured PDFs, {\it i.e.}, they are a consequence of the scattering process itself~\cite{Brodsky:2013oya,Brodsky:1989qz,Bauer}.  The $ I = 1$ Reggeon contribution to DDIS on the front-face nucleon then leads to flavor-dependent anti-shadowing~\cite{Brodsky:1989qz,Brodsky:2004qa}.  This could explain why the NuTeV charged current measurement $\mu A \to \nu X$ scattering does not appear to show anti-shadowing, in contrast to deep inelastic electron-nucleus scattering as discussed in  Ref. ~\cite{Schienbein:2007fs} and illustrated in Fig.\ref{Comparison}.



Note that final-state interactions 
can give non-trivial contributions to other classes of deep inelastic scattering processes at leading twist and thus survive at high $Q^2$ and high $W^2 = (q+p)^2.$  For example, the pseudo-$T$-odd Sivers effect~\cite{Brodsky:2002cx} is directly sensitive to the rescattering of the struck quark. 
\footnote{The ``handbag" approximation to deeply virtual Compton scattering  defines the ``static"  contribution~\cite{Brodsky:2008xe,Brodsky:2009dv} to the measured PDFs, Transverse Momentum Distributions (TMDs), etc..}  
Similarly,  DDIS  involves the exchange of gluons after the quark has been struck by the lepton~\cite{Brodsky:2002ue}.  In each case the corresponding scattering amplitude is not given by the handbag diagram since interactions between the two currents, $J^\mu(x)$ and $J^\nu(y)$, with vanishing invariant separation $(x-y)^2 \sim {1\over Q^2}$, are essential.
The FSI associated with the Sivers effect (``lensing" corrections \cite{Brodsky:2002cx,Burkardt:2003uw}) survive when both $W^2$ and $Q^2$ are large since the vector gluon couplings grow with energy.  However, in this case part of the final state phase can be associated with a Wilson line as an augmented Light Front Wave Function (LFWF) ~\cite{Brodsky:2010vs} which does not affect the $x_{Bj}$ moments, or the sum rules.  

Finally, analyzing shadowing in the dipole formalism, which uses the target rest frame, two scales are important. One is the coherence length (inverse of the momentum transfer), which has to be much larger than the nucleon separation. The other is the $\bar{q} q $ transverse separation, which has to be larger enough so that the photon interacts with a sizable cross section. This is related to the color transparency of small  transfers size fluctuations. These extra scales, which can be rather large compared to typical QCD scales, put strong doubts into the application of the OPE in nuclear shadowing processes, and therefore on the validity of the nuclear momentum sum rule.

\vspace{1cm}
In conclusion, we have illustrated why anti-shadowing of nuclear structure functions is non-universal, {\it i.e.},  flavor dependent, and why shadowing and anti-shadowing phenomena are incompatible with the application of OPE.  As a consequence, sum rules do not apply to  nuclear parton distribution functions. 

Even in the case of the proton, Mueller \cite{Mueller} has noted that the OPE applied to DIS fails at small $x_{Bj}$. The mechanism is however different: due to  the diffusion of gluons to small values of momentum transfer it is not possible to separate soft and hard scales in this region. This might be related to the fact that at small $x$ there are Fock states with a large number of small x partons, due to processes such as parton fusion and overlapping, which means that the invariant mass of these configurations is undefined. Therefore the usual derivation of sum rules based on the handbag diagram for (forward) double virtual Compton scattering may be inapplicable even on a single nucleon.

\section*{Acknowledgments}
This research was supported by the Department of Energy  contracts DE--AC02--76SF00515 (SJB) and DE-SC0016286 (SL).  
SLAC-PUB-XXXXX, by CONICYT (Chile) under Grants No. 1180232 and PIA/Basal FB0821.

\end{document}